\documentclass[twocolumn,preprintnumbers,amsmath,amssymb]{revtex4}
\usepackage{dcolumn}
\usepackage{bm}
\usepackage{graphicx}
\usepackage{amsmath}
\begin{document}

\title{Analytical results for a spin-orbit coupled atom held in a non-Hermitian double well under synchronous combined modulations}
\author{Xin Xie, \ Jiaxi Cui, \ Zhida Luo, \ Yuqiong Xie, \ Wenjuan Li, \ Wenhua Hai, and Yunrong Luo\footnote{Corresponding author: lyr\underline{ }1982@hunnu.edu.cn}}
\affiliation{Key Laboratory of Low-dimensional Quantum Structures and Quantum Control of Ministry
of Education, and Key Laboratory for Matter Microstructure and Function of Hunan Province, School of Physics and Electronics, Hunan Normal University, Changsha 410081, China}

\begin{abstract}
We propose a simple method of synchronous combined modulations to generate the exact analytic solutions for a spin-orbit (SO) coupled ultracold atom held in a non-Hermitian double-well potential. Based on the obtained analytical solutions, we mainly study the parity-time ($\mathcal{PT}$) symmetry of this system and the system stability for both balanced and unbalanced gain-loss between two wells. Under balanced gain and loss, the effect of the proportional constants between synchronous combined modulations and the SO-coupling strength on the $\mathcal{PT}$-symmetry breaking is revealed analytically. Surprisingly, we find when the Zeeman field is present, the stable spin-flipping tunneling between two wells can not occur in the non-Hermitian SO-coupled ultracold atomic system, but the stable spin-conserving tunneling can be performed. Under unbalanced gain and loss, the unique set of parameter conditions that can cause the system to stabilize is found. The results may provide a possibility for the exact control of $\mathcal{PT}$-symmetry breaking and quantum spin dynamics in a non-Hermitian SO-coupled system.

\end{abstract}

\maketitle

\section{Introduction}

The search for analytically exact solutions to quantum systems has always been a very challenging task in physics\cite{barnes109}. Due to the universal fact that it is extremely hard to acquire the exact solutions of time-dependent Schr\"{o}dinger equation except for a few specific situations. As is well-known, the analytical results play a crucial role in revealing the nature of physical phenomena. Not only that, they also provide many advantages in developing analytic methods to manipulate quantum states\cite{vion296, cole410, wu98} and qubit control\cite{economou, greilich, poem, nakamura398}. Specially, such exact analytical results are conducive to the development of manipulations that are both accurate and robust without the requirement for long control sequences\cite{motzoi, chow, gambetta, econo}. However, it is poignant that most of the exact solutions obtained are in two-level systems\cite{Zakrzewski32, rosen40, rabi51, hai87, luoxb95, JC51}, and the solutions are expressed by complicated especial functions\cite{bambini, hioe30, robinson, hioe32, ishkh, vitanov9, gango, bezver, sime}, for instance, Heun function\cite{jha81, jha82, xie82, ishk47, zhang93}, the Weber function\cite{landau2, zener137}, and so on. The accurate solutions of the simple form are very rare\cite{rabi51, hai87, luoxb95}, especially for multi-level systems.

In the recent years, the non-Hermitian physics has attracted much attention of almost all areas of physics\cite{moi2011, ashida, bergholtz, xu65}, including atomic and molecular physics\cite{lee2014}, mesoscopic physics\cite{rotter, rudner}, optomechanical systems\cite{xu2016}, spin and magnetic systems\cite{giorgi, galda}, etc. Many of exotic phenomena have been widely discovered both theoretically and experimentally, for instance, the exceptional topology\cite{bergholtz}, the localization transitions\cite{hatano}, the non-Hermitian bulk-boundary correspondence\cite{lee116}, the non-Hermitian skin effect\cite{gongjb}, and the spontaneous parity-time ($\mathcal{PT}$) symmetry breaking\cite{della87, luo110, gong2015, yang94, chit119, li2019, xiao2017}. In quantum mechanics, the requirement of a Hermitian Hamiltonian guarantees the existence of a real spectrum and the conservation of total probability. However, for a non-Hermitian Hamiltonian, the system can possess a complex energy spectrum whose imaginary part may be positive or negative. The positive imaginary part describes the probability increase exponentially that means the system is unstable. The negative imaginative part describes the probability decrease that can be used to stimulate decay phenomena\cite{okolo}. Particularly, when the non-Hermitian Hamiltonian of the system is parity and time-reversal symmetric, that is, $\mathcal{PT}$-symmetric, the purely real energy spectrum can still existence\cite{bender80, bender89}. An important characteristic of the $\mathcal{PT}$-symmetric system is the spontaneous $\mathcal{PT}$-symmetry breaking, where the energy spectrum changes from real (exact phase) to complex (broken phase) when the gain-loss coefficient $\vartheta$ exceeds a critical threshold $\vartheta_{\mathcal{PT}}$. The transition point $\vartheta=\vartheta_{\mathcal{PT}}$ is called the exceptional point (EP)\cite{miri}, which plays a key role in many $\mathcal{PT}$ studies\cite{makris, longhi103, feng333, regen488, hodaei, fleury, liu117, luoxb2023}.

In condensed matter physics, spin-orbit (SO) coupling is the interaction between spin of a particle and its motion, which plays a significant role in some interesting physical phenomena, for instance, spin-Hall effect\cite{kato} and topological insulators\cite{bernevig}. In recent years, the experimental realization of artificial SO-coupling of ultracold atoms has provided a new platform for studying the quantum dynamics of ultracold atomic systems\cite{lin471, wang109, cheuk109, zhang109, huang12, wu354, zhang128}. Many of novel quantum spin dynamical phenomena for SO-coupled ultracold atomic systems have been discovered, for instance, selective spin transport in a double-well potential\cite{yu90}, spin Josephson effects\cite{zhang609, garcia89, citro224}, dynamical suppression of tunneling\cite{kart97}, transparent control of spin dynamics\cite{luo39}, coherent control of spin-dependent localization\cite{luo93}, spin tunneling dynamics\cite{luo103}, and so on. However, all of these works have been investigated in the well-isolated Hermitian systems, much less is known about the role dissipation and SO-coupling play in the $\mathcal{PT}$-symmetry breaking and stability of non-Hermitian SO-coupled ultracold atomic systems\cite{luo22, ren, luo55}, which is particularly expected to be analytically revealed by applying exact solutions.

In this paper, we have presented a set of simple exact solutions for a single SO-coupled ultracold atom confined in a non-Hermitian double well under synchronous combined modulations. Based on the obtained exact solutions, we find when the monochromatically periodic modulation has a zero static component, the system is always stable regardless of the values of other parameters. When the periodic modulation has a nonzero static component, the $\mathcal{PT}$-symmetry of the system and the system stability are investigated analytically for both balanced and unbalanced gain-loss between two wells. Under balanced gain and loss, for the absence of Zeeman field, the $\mathcal{PT}$-symmetry breaking only depends on the proportional constant $\beta$ between the time-dependent gain-loss coefficient and tunneling rate. However, for the presence of Zeeman field, the $\mathcal{PT}$-symmetry breaking depends on the competition between the proportional constant $\beta$ and the SO-coupling strength $\gamma$. Not only that, we also surprisingly find that the stable spin-flipping tunneling between two wells can not occur in this case for the non-Hermitian SO-coupled ultracold atomic system. Under unbalanced gain and loss, the only set of parameter conditions that can cause system stability is found. The results may be useful for the coherent control of $\mathcal{PT}$-symmetry breaking and the stability of spin dynamics in a non-Hermitian SO-coupled system.

\section{EXACT ANALYTIC SOLUTIONS UNDER SYNCHRONOUS COMBINED MODULATIONS AND THEIR APPLICATIONS}

\subsection{Model System}

We consider a single SO-coupled ultracold atom confined in a driven non-Hermitian double well and the system can be described by a non-Hermitian Hamiltonian \cite{luoxb95, luo22, luo2023, zouml}
\begin{eqnarray}\label{eq1}
\hat{H}(t)&=& -\nu(t) (\hat{a}_{l}^{\dag}e^{-i \pi \gamma\hat{\sigma}_{y}}\hat{a}_{r}+H.c.) + \frac{\Omega(t)}{2} \sum_{j} (n_{j \uparrow}-n_{j \downarrow}) \nonumber\\ &+& i \sum_{\sigma} [\varepsilon_{l}(t)n_{l \sigma}-\varepsilon_{r}(t)n_{r \sigma}].
\end{eqnarray}
Here $\hat{a}_{j}^{\dag}=(\hat{a}_{j \uparrow}^{\dag},\hat{a}_{j \downarrow}^{\dag})$ and $\hat{a}_{j}=(\hat{a}_{j \uparrow},\hat{a}_{j \downarrow})^{T}$ ($T$ denotes the matrix transpose). $\hat{a}_{j \sigma}^{\dag}$ ($\hat{a}_{j \sigma}$) is the creation (annihilation) operator of a pseudospin-$\sigma$ ($\sigma=\uparrow, \downarrow$) atom in the $j$th $(j=l,r)$ well. \emph{H.c.} represents the Hermitian conjugate of the preceding term. $\hat{n}_{j\sigma}=\hat{a}_{j \sigma}^{\dag}\hat{a}_{j \sigma}$ denotes the number operator for spin $\sigma$ in the $j$th well, $\gamma$ denotes the effective SO-coupling strength, and $\hat{\sigma}_{y}$ is the $y$ component of Pauli operator. $\nu(t)$ is the time-dependent tunneling rate without SO-coupling\cite{luo2023}, $\Omega(t)$ is the time-dependent Zeeman field\cite{luo103}, and $\varepsilon_j(t)$ denotes the time-dependent gain-loss coefficient\cite{zouml, kreibich90, kreibich87}.

Taking the Fock states $|0, \sigma\rangle$ and $|\sigma, 0\rangle$ as basis vectors, the quantum state of the SO-coupled ultracold atomic system can be expanded as
\begin{eqnarray}\label{eq2}
|\psi(t)\rangle &=& a_1(t)|0,\uparrow\rangle + a_2(t)|0,\downarrow\rangle + a_3(t)|\uparrow,0\rangle \nonumber\\ &+& a_4(t)|\downarrow,0\rangle,
\end{eqnarray}
where $|0, \sigma\rangle$ ($|\sigma, 0\rangle$) represents the quantum state of a spin-$\sigma$ atom occupying the right (left) well and no atom in the left (right) well, and $a_{m}(t)$ ($m=1, 2, 3, 4$) denotes the time-dependent probability amplitude, for instance, $a_{1}(t)$ denotes the time-dependent probability amplitude of the spin atom being in Fock state $|0, \uparrow\rangle$. The corresponding probability reads $P_{m}(t)=|a_{m}(t)|^2$.
Inserting equations (1) and (2) into Schr\"{o}dinger equation $i\frac{\partial|\psi(t)\rangle}{\partial t}=\hat{H}(t)|\psi(t)\rangle$
results in the coupled equations
\begin{eqnarray}\label{eq3}
i\dot{a}_1(t)&=&-\nu(t) \cos(\pi \gamma) a_3(t)-\nu(t) \sin(\pi \gamma) a_4(t)\nonumber\\&+&[\frac{\Omega(t)}{2}-i\varepsilon_r(t)] a_1(t),\nonumber\\
i\dot{a}_2(t)&=&-\nu(t) \cos(\pi \gamma) a_4(t)+\nu(t) \sin(\pi \gamma) a_3(t)\nonumber\\&+&[-\frac{\Omega(t)}{2}-i\varepsilon_r(t)] a_2(t),\nonumber\\
i\dot{a}_3(t)&=&-\nu(t) \cos(\pi \gamma) a_1(t)+\nu(t) \sin(\pi \gamma) a_2(t)\nonumber\\&+&[\frac{\Omega(t)}{2}+i\varepsilon_l(t)] a_3(t),\nonumber\\
i\dot{a}_4(t)&=&-\nu(t) \cos(\pi \gamma) a_2(t)-\nu(t) \sin(\pi \gamma) a_1(t)\nonumber\\&+&[-\frac{\Omega(t)}{2}+i\varepsilon_l(t)] a_4(t).
\end{eqnarray}

We can see that it is very difficult to solve exactly equation (3), because of the time-dependent coefficients. Here, we will attempt to obtain the exact solutions of equation (3) under synchronous combined modulations.
For the synchronous combined modulations, we mean that the time-dependent driving functions $\Omega(t)$, $\varepsilon_j(t)$ and $\nu(t)$ obey the relations $\Omega(t)=\alpha\nu(t)$, $\varepsilon_l(t) = \kappa\nu(t)$, and $\varepsilon_r(t) = \beta\nu(t)$, in which $\alpha$, $\kappa$ and $\beta$ are proportional constants and we assume they are non-negative numbers. In this case, in order to obtain the exact solutions of equation (3), we introduce a new time variable $\tau=\tau(t)=\int \nu(t)dt$, then, equation (3) reduces to
\begin{eqnarray}\label{eq4}
i\frac{d{a}_1(t)}{d\tau}&=&- \cos(\pi \gamma) a_3(t)- \sin(\pi \gamma) a_4(t)\nonumber\\&+&[\frac{\alpha}{2}-i\beta] a_1(t),\nonumber\\
i\frac{d{a}_2(t)}{d\tau}&=&- \cos(\pi \gamma) a_4(t)+ \sin(\pi \gamma) a_3(t)\nonumber\\&+&[-\frac{\alpha}{2}-i\beta] a_2(t),\nonumber\\
i\frac{d{a}_3(t)}{d\tau}&=&- \cos(\pi \gamma) a_1(t)+ \sin(\pi \gamma) a_2(t)\nonumber\\&+&[\frac{\alpha}{2}+i\kappa] a_3(t),\nonumber\\
i\frac{d{a}_4(t)}{d\tau}&=&- \cos(\pi \gamma) a_2(t)- \sin(\pi \gamma) a_1(t)\nonumber\\&+&[-\frac{\alpha}{2}+i\kappa] a_4(t),
\end{eqnarray}
which is the linear differential equations of $\tau$ with constant coefficients and it can be solved precisely. We introduce the stationary-like solutions of equation (4) as $a_1(t) =
Ae^{-i \lambda\tau}$, $a_2(t) = Be^{-i \lambda\tau}$, $a_3(t) = Ce^{-i \lambda\tau}$, and $a_4(t) =De^{-i \lambda\tau}$, where $A$, $B$, $C$, and $D$ are constants, and $\lambda $ is the characteristic value. Inserting them into equation (4), we can obtain four characteristic values $\lambda_m$ and the corresponding constants $A_m$, $B_m$, $C_m$, $D_m$ for $ m= 1,2,3,4$ as
\begin{eqnarray}\label{eq5}
B_{1(2)}&=& A_{1(2)}\cot(\pi \gamma)-\frac{A_{1(2)}}{2}\csc(2\pi \gamma)[\frac{\eta}{2\alpha}\nonumber\\&-&i(\beta+\kappa)][\alpha+i(\beta+\kappa)\mp\sqrt{\zeta-\eta}], \nonumber\\
B_{3(4)}&=& A_{3(4)}\cot(\pi \gamma)+\frac{A_{3(4)}}{2}\csc(2\pi \gamma)[\frac{\eta}{2\alpha}\nonumber\\&+&i(\beta+\kappa)][\alpha+i(\beta+\kappa)\mp\sqrt{\zeta+\eta}], \nonumber\\
C_{1(2)}&=& \frac{A_{1(2)}}{2} \sec(\pi \gamma)[-i(\beta+\kappa)+\frac{\eta}{2\alpha}],\nonumber\\
C_{3(4)}&=&\frac{A_{3(4)}}{2} \sec(\pi \gamma)[-i(\beta+\kappa)-\frac{\eta}{2\alpha}],\nonumber\\
D_{1(2)}&=&\frac{A_{1(2)}}{2}\csc(\pi \gamma)[\alpha-\frac{\eta}{2\alpha}\mp\sqrt{\zeta-\eta}],\nonumber\\
D_{3(4)}&=&\frac{A_{3(4)}}{2}\csc(\pi \gamma)[\alpha+\frac{\eta}{2\alpha}\mp\sqrt{\zeta+\eta}],\nonumber\\
\lambda_{1(2)}&=&-\frac{1}{2}i(\beta-\kappa)\pm\frac{\sqrt{\zeta-\eta}}{2},\nonumber\\
\lambda_{3(4)}&=&-\frac{1}{2}i(\beta-\kappa)\pm\frac{\sqrt{\zeta+\eta}}{2}.
\end{eqnarray}
Here the constants $\eta=2\alpha\sqrt{-(\beta+\kappa)^2+4\cos(\pi \gamma)^2}$ and $\zeta=4+\alpha^2-(\beta+\kappa)^2$. Such that the four accurate stationary-like states of the system can be obtained as
\begin{eqnarray}\label{eq6}
|\psi_{m}(t)\rangle&=&(A_{m}|0,\uparrow\rangle+B_{m}|0,\downarrow\rangle + C_{m}|\uparrow,0\rangle + D_{m}|\downarrow,0\rangle)\nonumber\\ &\times& e^{-i\lambda_{m}\tau}
\end{eqnarray}
for $m=1,2,3,4$. According to the principle of linear superposition of quantum mechanics, we can readily obtain the general exact solution of equation (4) by employing equation (6) to the linear superposition
as
\begin{eqnarray}\label{eq7}
|\psi(t)\rangle = \sum^{4}_{m=1} s_m|\psi_m (t)\rangle,
\end{eqnarray}
where $s_m$ is an arbitrary superposition coefficient. Compared equation (7) with equation (2), the probability amplitudes are renormalized as
\begin{eqnarray}\label{eq7}
a_1(t)&=&\sum^{4}_{m=1}s_mA_me^{-i\lambda_m \tau},
a_2(t) = \sum^{4}_{m=1}s_mB_me^{-i\lambda_m \tau},\nonumber\\
a_3(t)&=&\sum^{4}_{m=1}s_mC_me^{-i\lambda_m \tau},
a_4(t) = \sum^{4}_{m=1}s_mD_me^{-i\lambda_m \tau},\nonumber\\
\end{eqnarray}
with the undetermined constant $s_m A_m$ being determined by the initial condition, which are the general exact solutions of equation (3) under the synchronous combined modulations. Obviously, for any function $\nu(t)$, there will be an exact analytical solution to the time-dependent Schr\"{o}dinger equation. Therefore, it enables us to produce an infinite variety of general exact solution (8) for the non-Hermitian SO-coupled ultracold atomic system. Here, we will select a monochromatically periodic modulation as an example to study the $\mathcal{PT}$-symmetry and stability of the system by applying the exact analytic solutions (5)-(8) under the synchronous combined modulations.

\subsection{Synchronous combined modulation of a monochromatically periodic driving form}

In this paper, we focus on the monochromatically periodic driving form
\begin{eqnarray}\label{eq9}
\nu(t) = \nu_1+\nu_2 \cos \omega t,
\end{eqnarray}
such that the new time variable is given by $\tau=\int \nu(t)dt=\nu_1 t+\frac{\nu_2}{\omega}\sin \omega t$. In this situation, the stationary-like state in equation (6) can be rewritten by
\begin{eqnarray}\label{eq10}
|\psi_{m}(t)\rangle=|\varphi_{m}(t)\rangle e^{-iE_m t}
\end{eqnarray}
with
\begin{eqnarray}\label{eq11}
|\varphi_{m}(t)\rangle&=&(A_{m}|0,\uparrow\rangle+B_{m}|0,\downarrow\rangle + C_{m}|\uparrow,0\rangle + D_{m}|\downarrow,0\rangle)\nonumber\\ &\times& e^{-i\frac{\lambda_{m}\nu_2}{\omega}\sin \omega t}
\end{eqnarray}
and
\begin{eqnarray}\label{eq12}
E_m=\lambda_{m}\nu_1.
\end{eqnarray}
According to the Floquet theorem\cite{shirley, sambe}, the state $|\varphi_{m}(t)\rangle$ is called Floquet state and $E_m$ is the Floquet quasienergy. For a non-Hermitian system, the quasienergy $E_m$ has complex value.
The imaginary part Im$(E_m)$ of the complex quasienergy $E_m$ plays a crucial role in the system stability. Based on the stability analysis\cite{luo22}, we know that the system is stable in two cases: (i) when all of Im$(E_m)$ are equal to zero, that is to say, the quasienergies are all real, the time-evolutions of probabilities are periodic and the system is stable; (ii) when some of Im$(E_m)$ are equal to zero and the others are less than zero, the time-evolutions of probabilities tend to be constant at time $t\rightarrow\infty$ and the system is also stable. Specially, it is noticed that when the constant modulation $\nu_1=0$, this will lead to the quasienergy $E_m=0$, which means the system is always stable regardless of the values of other parameters, e.g., see figure 1. In figure 1, the time-evolution curves of probabilities show the generalized Rabi oscillation\cite{gong2015}, so that the system is stable. The result is similar to that obtained in two-level system without SO-coupling in \cite{luoxb95}. However, in this work, we mainly study the situation of $\nu_1\neq0$.
\begin{figure}
\includegraphics[height=1.3in,width=2.2in]{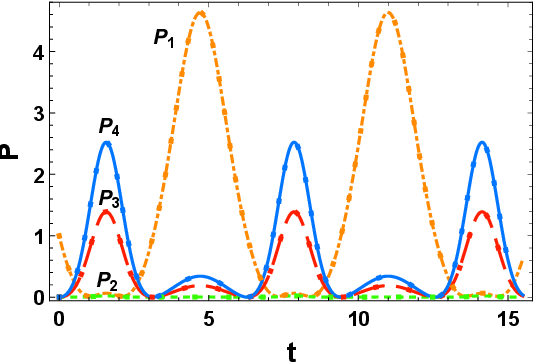}
\caption{\scriptsize{(Color online) Time evolutions of the probabilities versus the dimensionless time as $P_1=|a_1(t)|^2$ in state $|0,\uparrow\rangle$ (orange dashed-dotted curve), $P_2=|a_2(t)|^2$ in state $|0,\downarrow\rangle$ (green short-dashed curve), $P_3=|a_3(t)|^2$ in state $|\uparrow,0\rangle$ (red long-dashed curve), and $P_4=|a_4(t)|^2$ in state $|\downarrow,0\rangle$ (blue solid curve). The parameters are chosen as $\nu_1=0$, $\nu_2=1$, $\omega=1$, $\alpha=1$, $\beta=1$, $\kappa=2$, and $\gamma=0.3$, starting the system with a spin-up particle in the right well. Hereafter, the squares denote the
analytical results from equation (8), and the curves denote the numerical correspondences obtained from equation (3).}}
\end{figure}

\subsubsection{$\mathcal{PT}$-symmetry and system stability under balanced gain and loss}

Here, we first consider the situation of balanced gain and loss between two wells, namely, the loss (gain) coefficients of two wells take the same values, thus the proportional constants $\beta=\kappa$. In this situation, the Hamiltonian (1) is $\mathcal{PT}$-symmetric, due to $\hat{P}\hat{T}\hat{H}=\hat{H}\hat{P}\hat{T}$, where the parity operator and the time reversal operator are defined as $\hat{P}: l \leftrightarrow r$ and $\hat{T}: t \rightarrow -t, i \rightarrow -i$, respectively. In addition, from equation (5), we can infer
\begin{eqnarray}\label{eq13}
\lambda_1 = -\lambda_2 = \frac{\sqrt{\zeta-\eta}}{2}, \lambda_3 = -\lambda_4 = \frac{\sqrt{\zeta+\eta}}{2}
\end{eqnarray}
with $\zeta=4+\alpha^2-4\beta^2$ and $\eta=4\alpha\sqrt{-\beta^2+\cos(\pi \gamma)^2}$. It can be seen from equation (13) that when Im$(\sqrt{\zeta-\eta})=0$ holds, both Im$(\lambda_m)=0$ and Im$(E_m)=0$ will naturally hold, such that the system will be stable according to the above case (i) of stability analysis. Not only that, we can also see that when $\alpha=0$ (the absence of Zeeman field)\cite{luo55}, $\eta=0$ and $\zeta=4-4\beta^2$, thus, the critical threshold at which spontaneous $\mathcal{PT}$-symmetry breaking occurs is $\beta=1$. So clearly, the $\mathcal{PT}$-symmetry breaking only depends on the proportional constant $\beta$. When $\alpha\neq0$ (the presence of Zeeman field)\cite{yu90, luo22, luo2023}, the critical value of the $\mathcal{PT}$-symmetry breaking phase transition meets $\beta=|\cos \pi\gamma|$, therefore, the $\mathcal{PT}$-symmetry breaking depends on the competition between the proportional constant $\beta$ and the SO-coupling strength $\gamma$. In particular, it is surprisingly found that when $\gamma=n+0.5$ ($n=0,1,2,...$), the critical threshold $\beta=|\cos \pi\gamma|=0$. That is to say, the spontaneous $\mathcal{PT}$-symmetry breaking phase transition occurs for arbitrarily small proportional constant $\beta$.

Next, we will in detail analyze the effect of system parameters on the spontaneous $\mathcal{PT}$-symmetry breaking and the system stability as illustrated by figures 2-4, the boundaries between stable and unstable parameter regions are plotted by red curves, across which the transition of quasienergies from all real (unbroken $\mathcal{PT}$-symmetry) to complex (broken $\mathcal{PT}$-symmetry) will happen. In Figures 2(a)-(b), we plot the imaginary part Im$(\sqrt{\zeta-\eta})$ as a function of $\beta$ and $\gamma$ for different values of $\alpha$ with (a) $\alpha=0$ and (b) $\alpha=1$. It is obviously seen that the boundaries between stable and unstable parameter regions are in good agreement with the above analytic results. In figure 2(a), when the proportional constant $\beta$ is smaller than the critical threshold, $\beta=1$, a continuous stable parameter region exists, and this stable parameter region is independent of the SO-coupling strength $\gamma$. When the parameters are chosen in this stable region, the time evolutions of probabilities will show quasiperiodic oscillations with bound, see figure 2(c). In figure 2(c), the stable quasiperiodic population oscillations with spin-flipping between states $|0, \uparrow\rangle$ and $|\downarrow, 0\rangle$ occur. In figure 2(b), it can be seen that the stable parameter regions are discrete. The boundary between stable and unstable parameter regions satisfies $\beta=|\cos \pi\gamma|$. Specially, when $\gamma=n+0.5$ ($n=0,1,2,...$), it can be found from figure 2(b) that the stable parameter region vanishes in the non-Hermitian SO-coupled system. In other words, the stable spin-flipping tunneling between two wells can not occur for $\alpha\neq 0$ in the non-Hermitian SO-coupled system, e.g., see figure 2(d). In figure 2(d), the population oscillations with spin-flipping between states $|0, \uparrow\rangle$ and $|\downarrow, 0\rangle$ show exponential growth, thus, the system is unstable. However, when $\gamma=n$ ($n=0,1,2,...$), the stable parameter regions exist, which can be seen from figure 2(b). Such that the stable spin-conserving tunneling between two wells can happen (not shown here).

\begin{figure}[htp]\center
\includegraphics[height=1.3in,width=1.6in]{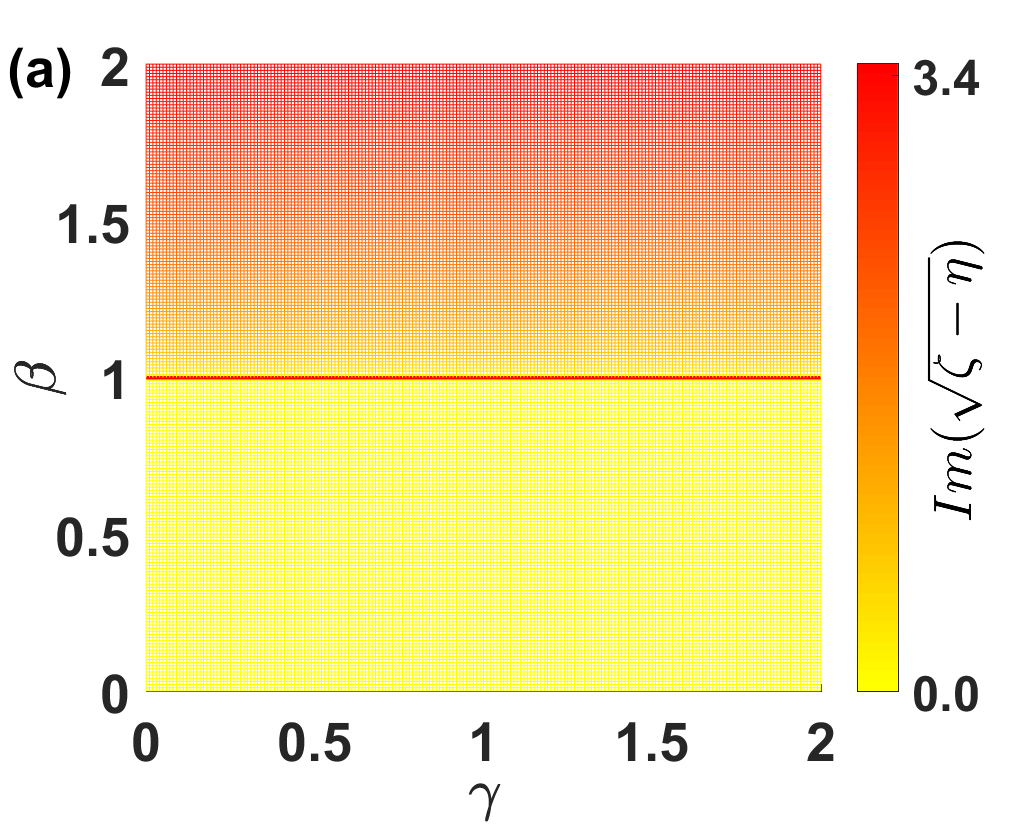}
\includegraphics[height=1.3in,width=1.6in]{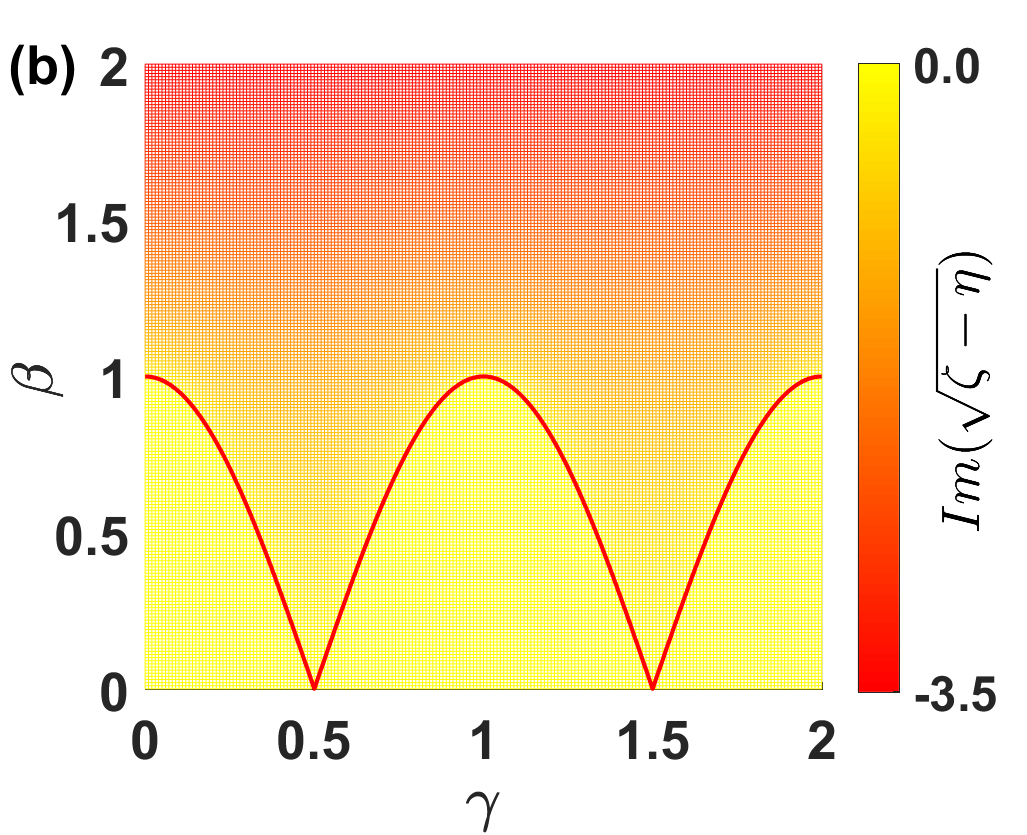}
\includegraphics[height=1.3in,width=1.6in]{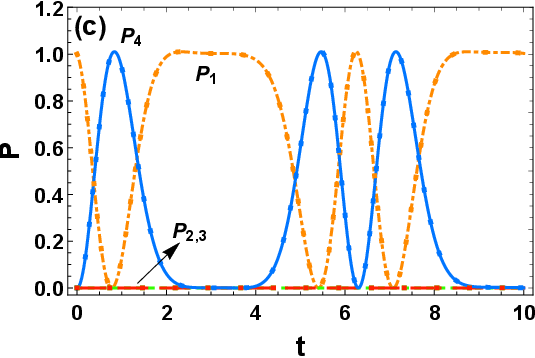}
\includegraphics[height=1.3in,width=1.6in]{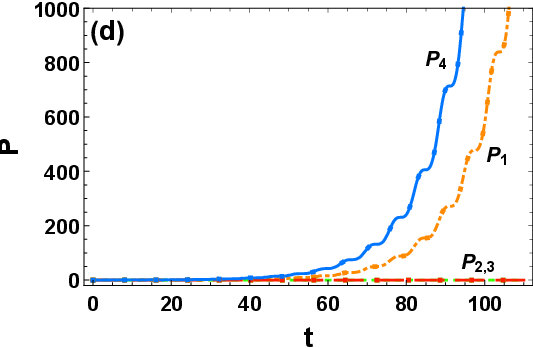}
\caption{\scriptsize{(Color online) Top panels depict the imaginary part Im$(\sqrt{\zeta-\eta})$ as a function of $\beta$ and $\gamma$ for (a) $\alpha=0$ and (b) $\alpha=1$. Hereafter, the red curves are the boundary between stable and unstable parameter regions. Bottom panels show the time-evolution curves of probabilities $P_{m}=|a_{m}(t)|^{2}$ for (c) $\alpha=0$ and (d) $\alpha=1$, starting the system with a spin-up particle in the right well. The other parameters are taken as $\beta=\kappa=0.1$, $\gamma=0.5$, $\nu_1=1$, $\nu_2=1$, and $\omega=1$.}}
\end{figure}

In figures 3(a)-(b), we plot Im$(\sqrt{\zeta-\eta})$ as a function of $\alpha$ and $\gamma$ for different proportional constants (a) $\beta=\kappa=0.3$ and (b) $\beta=\kappa=0.9$. The boundary between stable and unstable parameter regions satisfies $\gamma=n+\arccos(\pm \beta)/\pi$ ($n=0,1,2,...$). It is obviously seen that the width of the stable parameter regions becomes narrower with increasing $\beta$. The stable (or unstable) parameter regions are independent of the value of the proportional constant $\alpha$, except for the special value of $\alpha=0$. From figures 3(a)-(b), we find when the SO-coupling strength $\gamma=n+0.5$ ($n=0,1,2,...$), the system is unstable. Thus, the stable spin-flipping tunneling between two wells can not occur(not shown here). However, when $\gamma=n$ ($n=0,1,2,...$), the system is stable and the stable non-spin-flipping tunneling between states $|0, \uparrow\rangle$ and $|\uparrow, 0\rangle$ can be performed, see figure 3(c). In figure 3(d), we take a general value of $\gamma$ (e.g., $\gamma=0.1$) and $\alpha=1$ in the stable region of figure 3(b) to plot the time-evolution curves of probabilities. It can be seen from figure 3(d) that the quasiperiodic population oscillations between four Fock states happen.

\begin{figure}[htp]\center
\includegraphics[height=1.3in,width=1.6in]{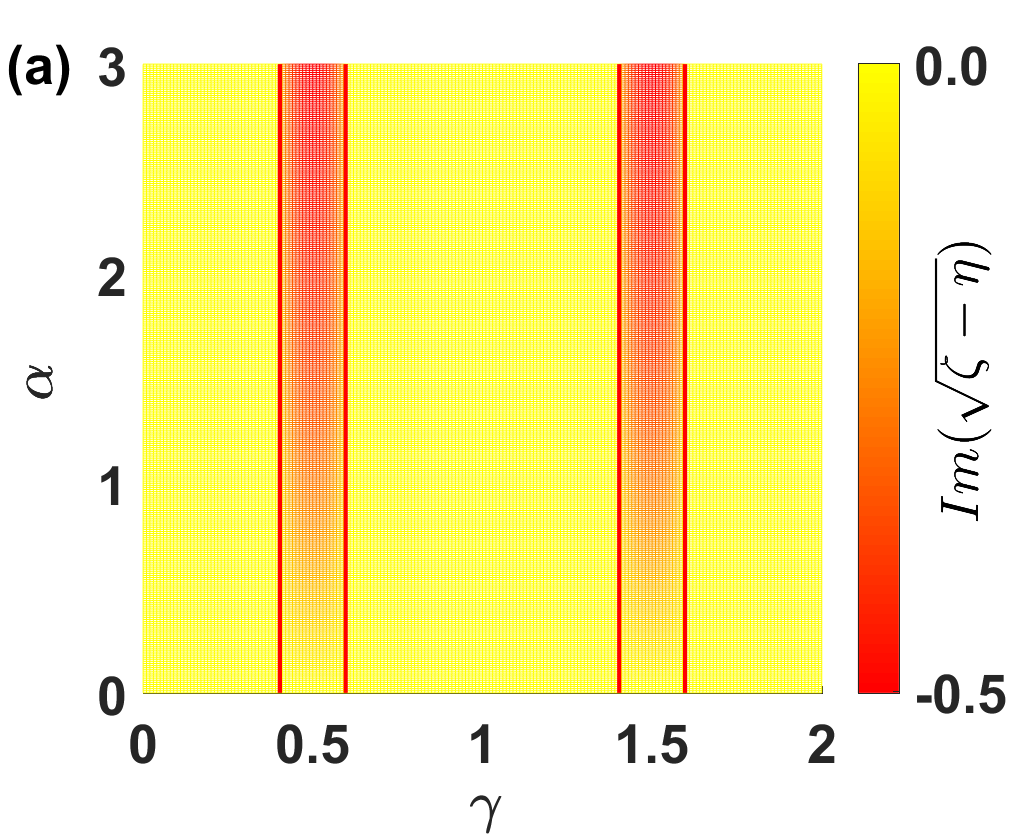}
\includegraphics[height=1.3in,width=1.6in]{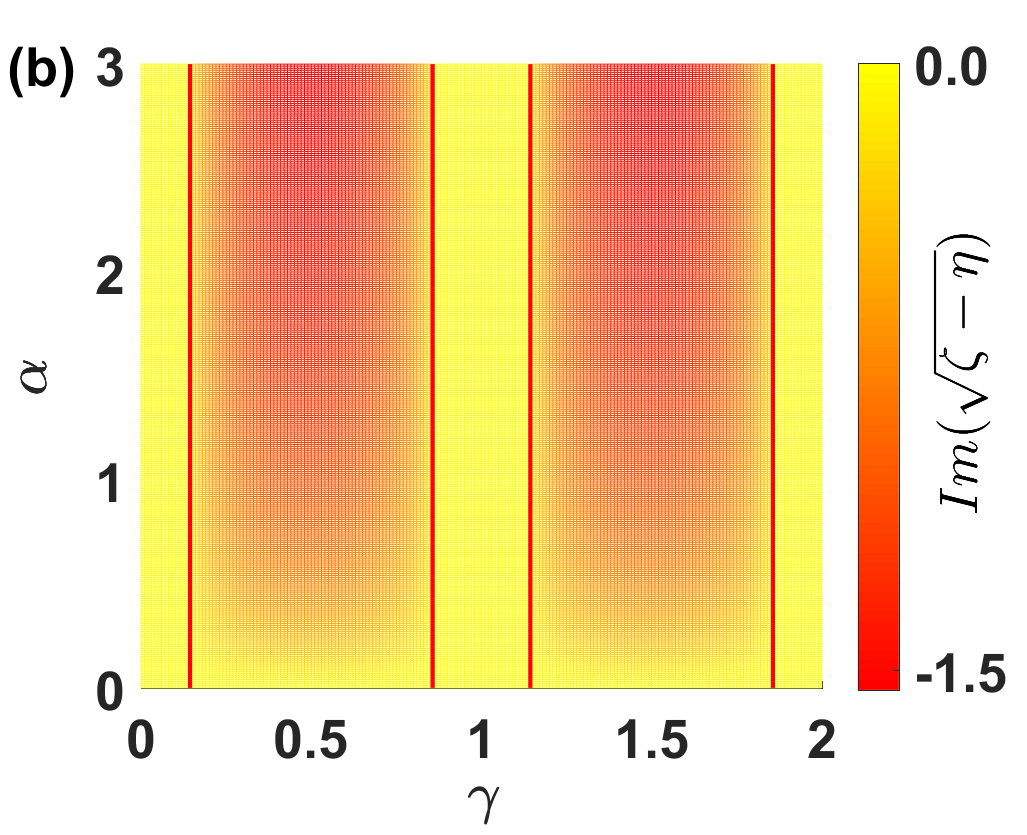}
\includegraphics[height=1.3in,width=1.6in]{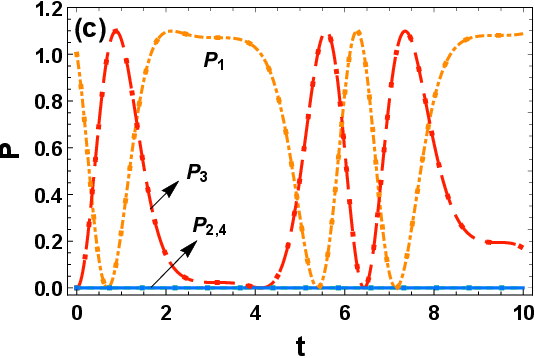}
\includegraphics[height=1.3in,width=1.6in]{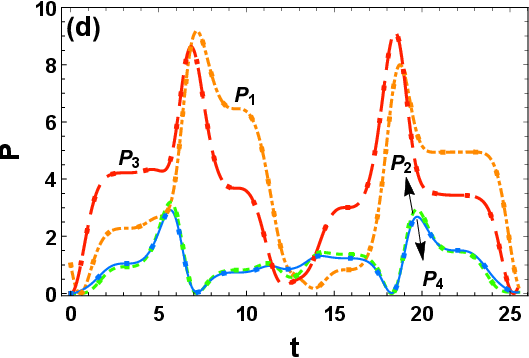}
\caption{\scriptsize{(Color online) Top panels depict the imaginary part Im$(\sqrt{\zeta-\eta})$ as a function of $\alpha$ and $\gamma$ for (a) $\beta=\kappa=0.3$ and (b) $\beta=\kappa=0.9$. Bottom panels show the time-evolution curves of probabilities $P_{m}=|a_{m}(t)|^{2}$ for (c) $\beta=\kappa=0.3$, $\gamma=1$; (d) $\beta=\kappa=0.9$, $\gamma=0.1$, starting the system with a spin-up particle in the right well. The other parameters are taken as $\alpha=1$, $\nu_1=1$, $\nu_2=1$, and $\omega=1$.}}
\end{figure}

In figures 4(a)-(b), we plot Im$(\sqrt{\zeta-\eta})$ as a function of $\alpha$ and $\beta$ for different SO-coupling strengths (a) $\gamma=\frac{1}{3}$ and (b) $\gamma=1$. Obviously, the boundaries between stable regions (unbroken $\mathcal{PT}$-symmetry) and unstable regions (broken $\mathcal{PT}$-symmetry) satisfy $\beta=|\cos \pi\gamma|=0.5$ in (a) and $\beta=|\cos \pi\gamma|=1$ in (b), respectively. In figure 4(c), we set the parameters $\alpha=1$ and $\beta=\kappa=0.1$ in the stable parameter region of figure 4(a) to plot the time-evolution curves of probabilities. It can be seen from figure 4(c) that the population oscillations with bound between four Fock states occur. In figure 4(d), we take the parameters $\alpha=1$ and $\beta=\kappa=1$ on the boundary between stable and unstable parameter regions (at the transition point or EP) to plot the time-evolution curves of probabilities. It is found that the probabilities $P_1$ and $P_3$ increase exponentially and the system is unstable, which is due to the EPs mechanism\cite{luoxb2023}.

\begin{figure}[htp]\center
\includegraphics[height=1.3in,width=1.6in]{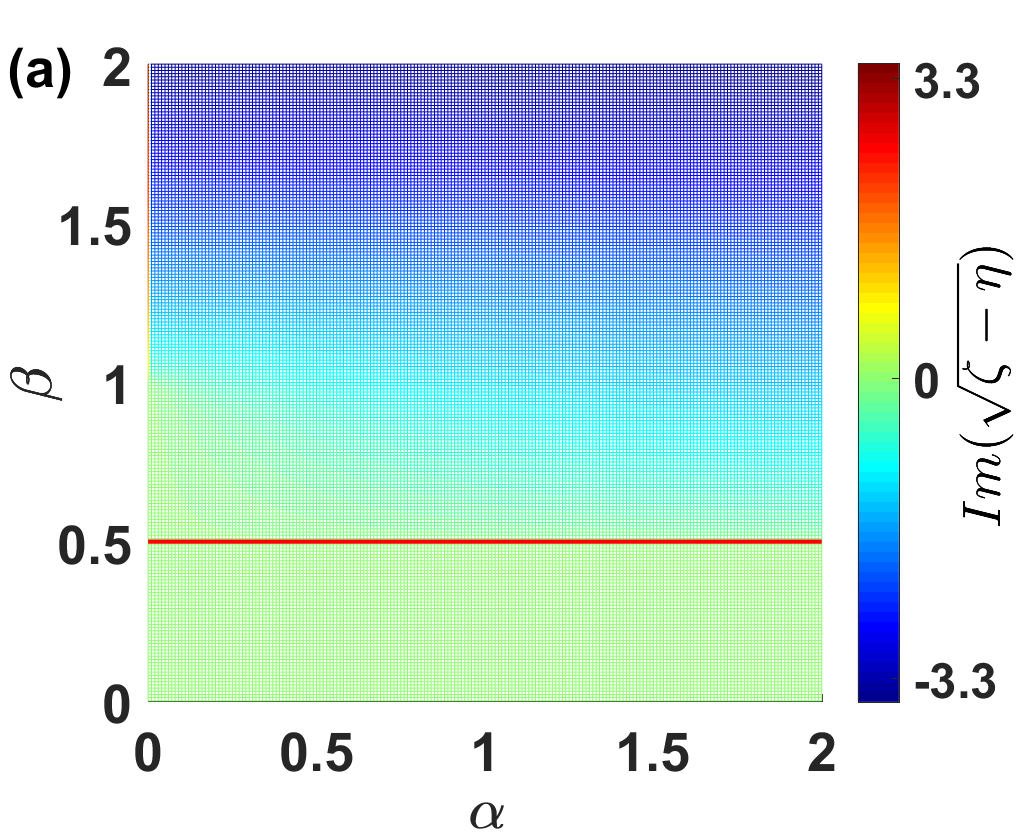}
\includegraphics[height=1.3in,width=1.6in]{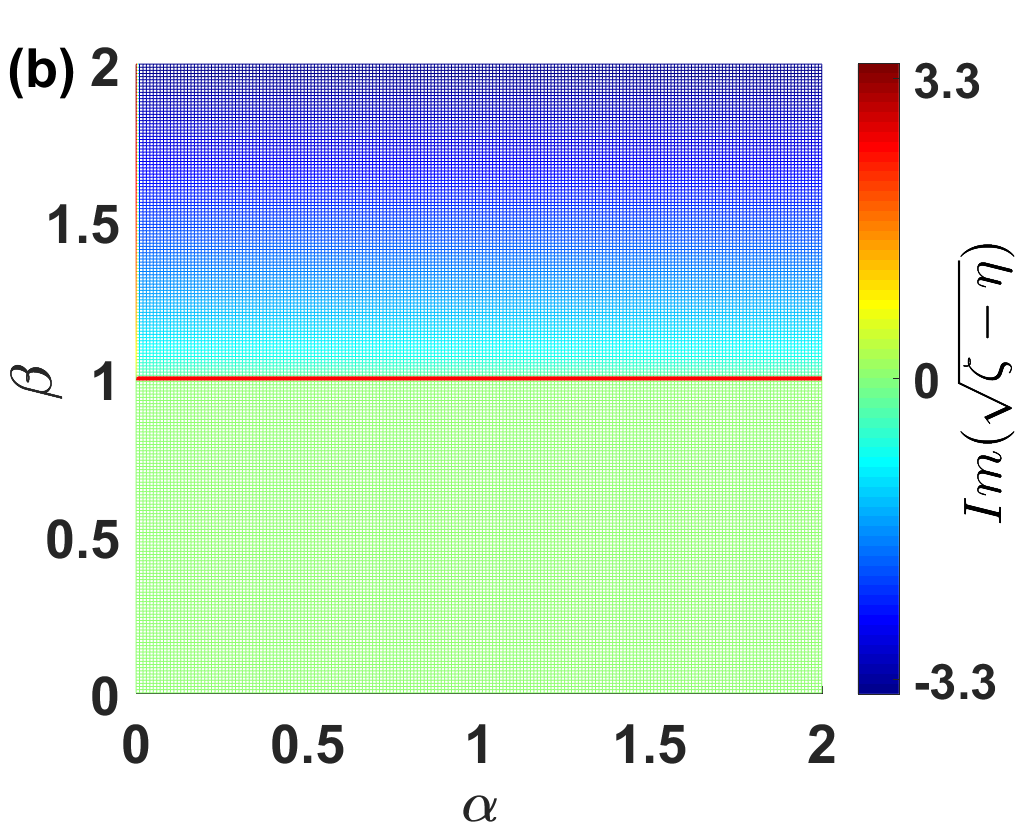}
\includegraphics[height=1.3in,width=1.6in]{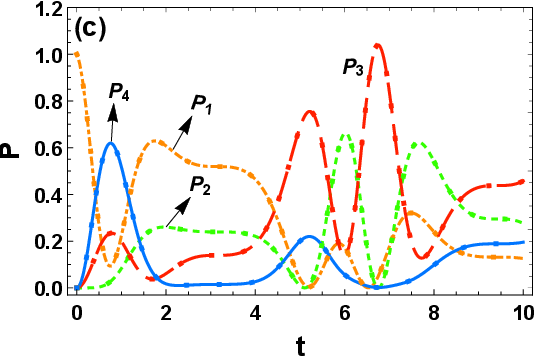}
\includegraphics[height=1.3in,width=1.6in]{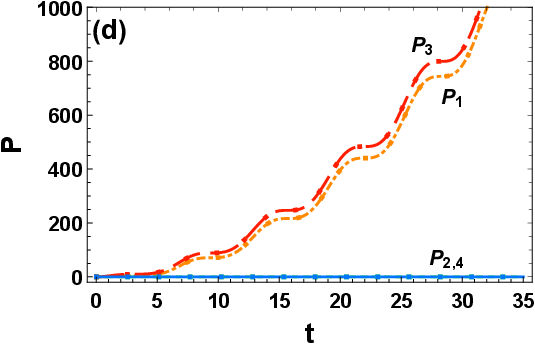}
\caption{\scriptsize{(Color online) Top panels depict the imaginary part Im$(\sqrt{\zeta-\eta})$ as a function of $\alpha$ and $\beta$ for (a) $\gamma=\frac{1}{3}$ and (b) $\gamma=1$. Bottom panels show the time-evolution curves of probabilities $P_{m}=|a_{m}(t)|^{2}$ for (c) $\beta=\kappa=0.1$, $\gamma=\frac{1}{3}$; (d) $\beta=\kappa=1$, $\gamma=1$, starting the system with a spin-up particle in the right well. The other parameters are taken as $\alpha=1$, $\nu_1=1$, $\nu_2=1$, and $\omega=1$.}}
\end{figure}

\subsubsection{System stability under unbalanced gain and loss}

\begin{figure}
\includegraphics[height=1.3in,width=1.6in]{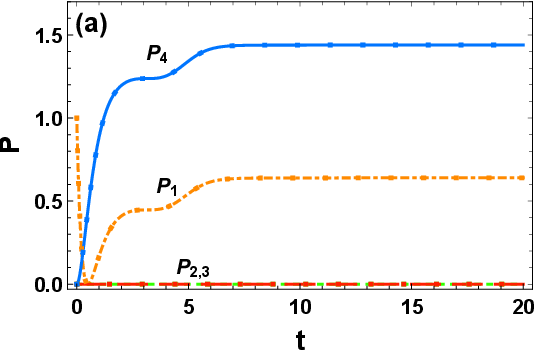}
\includegraphics[height=1.3in,width=1.6in]{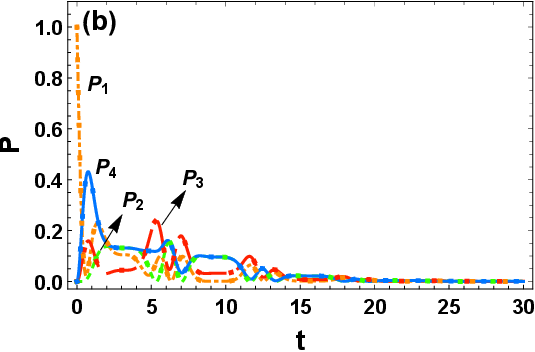}
\caption{\scriptsize{(Color online) Time-evolution curves of probabilities $P_{m}=|a_{m}(t)|^{2}$ for (a) $\alpha=0$, $\beta=\frac{3}{2}$, $\kappa=\frac{2}{3}$, $\gamma=0.5$; (b) $\alpha=1$, $\beta=0.7$, $\kappa=0.3$, $\gamma=\frac{1}{3}$, starting the system with a spin-up particle in the right well. The other parameters are taken as $\nu_1=1$, $\nu_2=1$, and $\omega=1$.}}
\end{figure}

Now we consider the unbalanced situation of gain and loss, namely, the proportional constants $\beta\neq\kappa$. To ensure the stability of the system, we find the proportional constant $\beta$ must be greater than $\kappa$\cite{luo22, luo55}, namely, $\beta>\kappa$. From equation (5), it is noticed that when the parameters satisfy
\begin{eqnarray}\label{eq14}
\eta=0,\ \beta-\kappa=\sqrt{-\zeta},
\end{eqnarray}
we can obtain
\begin{eqnarray}\label{eq15}
\lambda_1=\lambda_3=0,\ \lambda_2=\lambda_4=-i(\beta-\kappa).
\end{eqnarray}
According to the above case (ii) of stability analysis, the system is stable in this situation. Based on equation (14), we can derive the parameter conditions for the stability of the system as
\begin{eqnarray}\label{eq16}
\alpha=0,\ \beta*\kappa=1,\ \beta>\kappa,
\end{eqnarray}
under unbalanced gain and loss. As an example, we take the parameters $\alpha=0$, $\beta=\frac{3}{2}$, $\kappa=\frac{2}{3}$, $\gamma=0.5$, $\nu_1=1$, $\nu_2=1$, and $\omega=1$ obeying the stability conditions (16) to plot the time-evolution curves of probabilities for the particle initially occupying the state $|0, \uparrow\rangle$, as shown in figure 5(a). It can be seen that the probabilities $P_1(t)$ and $P_4(t)$ tend to stable values after a short period of time and $P_2(t)=P_3(t)=0$, which mean the system is stabilized. It is worth noting that if the system parameters cannot simultaneously satisfy the two conditions $\eta=0$ and $\beta-\kappa=\sqrt{-\zeta}$, the system will be unstable, e.g., see figure 5(b). In figure 5(b), the all probabilities decay to zero after a period of time and the spin particle is lost asymptotically. In addition, for the cases of $\eta>0$ and $\eta$ being complex, we can demonstrate analytically that no suitable parameters meet the above case (ii) of stability analysis (not shown here). Thus, for the situation of unbalanced gain and loss, the equation (16) is the unique set of parameter equilibrium conditions that can cause the system to stabilize.

\section{CONCLUSION AND DISCUSSION}

In summary, we have obtained the exact analytic solutions for a single SO-coupled ultracold atom confined in a non-Hermitian double-well potential under synchronous combined modulations. For the periodic modulation with a zero static component, the quasienergies of the system are all real and the system is always stable irrespective of the values of other parameters, which is similar to that obtained in two-level system without SO-coupling in \cite{luoxb95}. For the periodic driving with a nonzero static component, the $\mathcal{PT}$-symmetry of the system and the system stability have been revealed analytically for both balanced and unbalanced gain-loss between two wells. Under balanced gain and loss, the transition point (or EP) between unbroken $\mathcal{PT}$-symmetry (stable region) and broken $\mathcal{PT}$-symmetry (unstable region) is related to the presence or absence of Zeeman field. For the absence of Zeeman field ($\alpha=0$), the critical threshold of $\mathcal{PT}$-symmetry breaking is $\beta=1$. However, for the presence of Zeeman field ($\alpha\neq0$), the $\mathcal{PT}$-symmetry breaking depends on the competition between the proportional constant $\beta$ and the SO-coupling strength $\gamma$, and the crucial threshold is $\beta=|\cos \pi\gamma|$. Especially, it is surprisingly found that when the SO-coupling strength is half-odd, i.e., $\gamma=0.5,1.5,2.5,...$, the spontaneous $\mathcal{PT}$-symmetry breaking occurs for arbitrarily small proportional constant $\beta$ which extends the result obtained in the absence of Zeeman field in \cite{luo55}. Meanwhile, this also means that the stable spin-flipping tunneling between two wells can not occur in this situation for the non-Hermitian SO-coupled ultracold atomic system, see figure 2(d). But, the stable spin-conserving tunneling can be performed. Under unbalanced gain and loss, we have found that for the system to be stable, it has had to be in the absence of Zeeman field and the corresponding parameter conditions have been found, see equation (16), which is independent of the SO-coupling strength $\gamma$. Our results may find some applications in the coherent control of $\mathcal{PT}$-symmetry breaking and the system stability for a non-Hermitian SO-coupled system.

\section*{ACKNOWLEDGMENTS}

This work was supported by the Hunan Provincial Natural Science Foundation of China under Grants No. 2021JJ30435 and No. 2017JJ3208, the Scientific Research Foundation of Hunan Provincial Education Department under Grants No. 21B0063 and No. 18C0027, and the National Natural Science Foundation of China under Grant No. 11747034.


\begin{thebibliography}{999}

\bibitem{barnes109} Barnes E and Das Sarma S 2012 Analytically Solvable Driven Time-Dependent Two-Level Quantum Systems \emph{Phys. Rev. Lett.} 109 060401
\bibitem{vion296} Vion D, Aassime A, Cottet A, Joyez P, Pothier H, Urbina C, Esteve D and Devoret M H 2002 Manipulating the Quantum
State of an Electrical Circuit \emph{Science} 296 886
\bibitem{cole410} Cole B E, Williams J B, King B T, Sherwin M S and Stanley C R 2001 Coherent manipulation of semiconductor quantum bits with terahertz radiation \emph{Nature(London)} 410 60
\bibitem{wu98} Wu Y and Yang X 2007 Strong-Coupling Theory of Periodically Driven Two-Level Systems \emph{Phys. Rev. Lett.} 98 013601
\bibitem{economou} Economou S E, Sham L J, Wu Y and Steel D G 2006 Proposal for optical U(1) rotations of electron spin trapped in a quantum dot \emph{Phys. Rev. B} 74 205415
\bibitem{greilich} Greilich A, Economou S E, Spatzek S, Yakovlev D R, Reuter D, Wieck A D, Reinecke T L and Bayer M 2009 Ultrafast optical rotations of electron spins in quantum dots \emph{Nature Phys.} 5 262
\bibitem{poem} Poem E, Kenneth O, Kodriano Y, Benny Y, Khatsevich S, Avron J E and Gershoni D 2011 Optically Induced Rotation of an Exciton Spin in a Semiconductor Quantum Dot \emph{Phys. Rev. Lett.} 107 087401
\bibitem{nakamura398} Nakamura Y, Pashkin Yu A and Tsai J S 1999 Coherent control of macroscopic quantum states in a single-Cooper-pair box \emph{Nature(London)} 398 786
\bibitem{motzoi} Motzoi F, Gambetta J M, Rebentrost P and Wilhelm F K 2009 Simple Pulses for Elimination of Leakage in Weakly Nonlinear Qubits \emph{Phys. Rev. Lett.} 103 110501
\bibitem{chow} Chow J M, DiCarlo L, Gambetta J M, Motzoi F, Frunzio L, Girvin S M and Schoelkopf R J 2010 Optimized driving of superconducting artificial atoms for improved single-qubit gates \emph{Phys. Rev. A} 82 040305(R)
\bibitem{gambetta} Gambetta J M, Motzoi F, Merkel S T and Wilhelm F K 2011 Analytic control methods for high-fidelity unitary operations in a weakly nonlinear oscillator \emph{Phys. Rev. A} 83 012308
\bibitem{econo} Economou S E 2012 High-fidelity quantum gates via analytically solvable pulses \emph{Phys. Rev. B} 85 241401(R)
\bibitem{Zakrzewski32} Zakrzewski J 1985 Analytic solutions of the two-state problem for a class of chirped pulses \emph{Phys. Rev. A} 32 3748
\bibitem{rosen40} Rosen N and Zener C 1932 Double Stern-Gerlach Experiment and Related Collision Phenomena \emph{Phys. Rev.} 40 502
\bibitem{rabi51} Rabi I 1937 Space Quantization in a Gyrating Magnetic Field \emph{Phys. Rev.} 51 652
\bibitem{hai87} Hai W, Hai K and Chen Q 2013 Transparent control of an exactly solvable two-level system via combined modulations \emph{Phys. Rev. A} 87 023403
\bibitem{luoxb95} Luo X, Yang B, Zhang X, Li L and Yu X 2017 Analytical results for a parity-time-symmetric two-level system under synchronous combined modulations \emph{Phys. Rev. A} 95 052128
\bibitem{JC51} Jaynes E and Cummings F 1963 Comparison of quantum and semiclassical radiation theories with application to the beam maser \emph{Proc. IEEE} 51 89
\bibitem{bambini} Bambini A and Berman P R 1981 Analytic solutions to the two-state problem for a class of coupling potentials \emph{Phys. Rev. A} 23 2496
\bibitem{hioe30} Hioe F 1984 Solution of Bloch equations involving amplitude and frequency modulation \emph{Phys. Rev. A} 30 2100
\bibitem{robinson} Robinson E J 1985 Two-level systems driven by modulated pulses \emph{Phys. Rev. A} 31 3986
\bibitem{hioe32} Hioe F and Carroll C 1985 Two-state problems involving arbitrary amplitude and frequency modulations \emph{Phys. Rev. A} 32 1541
\bibitem{ishkh} Ishkhanyan A M 2000 New classes of analytic solutions of the two-level problem \emph{J. Phys. A} 33 5539
\bibitem{vitanov9} Vitanov N 2007 Complete population inversion by a phase jump: an exactly soluble model \emph{New J. Phys.} 9 58
\bibitem{gango} Gangopadhyay A, Dzero M and Galitski V 2010 Exact solution for quantum dynamics of a periodically driven two-level system \emph{Phys. Rev. B} 82 024303
\bibitem{bezver} Bezvershenkoa Y and Holod P 2011 Resonance in a driven two-level system: Analytical results without the rotating wave approximation \emph{Phys. Lett. A} 375 3936
\bibitem{sime} Simeonov L S and Vitanov N V 2014 Exactly solvable two-state quantum model for a pulse of hyperbolic-tangent shape \emph{Phys. Rev. A} 89 043411
\bibitem{jha81} Jha P K and Rostovtsev Y V 2010 Coherent excitation of a two-level atom driven by a far-off-resonant classical field: Analytical solutions \emph{Phys. Rev. A} 81 033827
\bibitem{jha82} Jha P K and Rostovtsev Y V 2010 Analytical solutions for a two-level system driven by a class of chirped pulse \emph{Phys. Rev. A} 82 015801
\bibitem{xie82} Xie Q and Hai W 2010 Analytical results for a monochromatically driven two-level system \emph{Phys. Rev. A} 82 032117
\bibitem{ishk47} Ishkhanyan A M and Grigoryan A E 2014 Fifteen classes of solutions of the quantum two-state problem in terms of the confluent Heun function \emph{J. Phys. A: Math. Theor.} 47 465205
\bibitem{zhang93} Zhang W, Jin K, Jin L and Xie X 2016 Analytic results for the population dynamics of a driven dipolar molecular system \emph{Phys. Rev. A} 93 043840
\bibitem{landau2} Landau L 1932 To the theory of energy transmission in collissions \emph{Phys. Zs. Sowjet.} 2 46
\bibitem{zener137} Zener C 1932 Non-Adiabatic Crossing of Energy Levels \emph{Proc. R. Soc. Lond. A} 137 696
\bibitem{moi2011} Moiseyev N 2011 \emph{Non-Hermitian Quantum Mechanics} (Cambridge: Cambridge University Press)
\bibitem{ashida} Ashida Y, Gong Z and Ueda M 2020 Non-Hermitian physics \emph{Adv. Phys.} 69 249
\bibitem{bergholtz} Bergholtz E J, Budich J C and Kunst F K 2021 Exceptional topology of non-Hermitian systems \emph{Rev. Mod. Phys.} 93 015005
\bibitem{xu65} Xu Z, Zhou Z, Cheng E, Lang L and Zhu S 2022 Gain/loss effects on spin-orbit coupled ultracold atoms in two-dimensional optical lattices \emph{Sci. China-Phys. Mech. Astron.} 65 283011
\bibitem{lee2014} Lee T E and Chan C-K 2014 Heralded Magnetism in Non-Hermitian Atomic Systems \emph{Phys. Rev. X} 4 041001
\bibitem{rotter} Rotter I and Bird J P 2015 A review of progress in the physics of open quantum systems: theory and experimen \emph{Rep. Prog. Phys.} 78 114001
\bibitem{rudner} Rudner M S and Levitov L S 2010 Phase transitions in dissipative quantum transport and mesoscopic nuclear spin pumping \emph{Phys. Rev. B} 82 155418
\bibitem{xu2016} Xu H, Mason D, Jiang L and Harris J G E 2016 Topological energy transfer in an optomechanical system with exceptional points \emph{Nature} 537 80
\bibitem{giorgi} Giorgi G-L 2010 Spontaneous $\mathcal{PT}$ symmetry breaking and quantum phase transitions in dimerized spin chains \emph{Phys. Rev. B} 82 052404
\bibitem{galda} Galda A and Vinokur V M 2016 Parity-time symmetry breaking in magnetic systems \emph{Phys. Rev. B} 94 020408
\bibitem{hatano} Hatano N and Nelson D R 1996 Localization Transitions in Non-Hermitian Quantum Mechanics \emph{Phys. Rev. Lett.} 77 570
\bibitem{lee116} Lee T E 2016 Anomalous Edge State in a Non-Hermitian Lattice \emph{Phys. Rev. Lett.} 116 133903
\bibitem{gongjb} Li L, Lee C H and Gong J 2020 Topological Switch for Non-Hermitian Skin Effect in Cold-Atom Systems with Loss \emph{Phys. Rev. Lett.} 124 250402
\bibitem{della87} Della Valle G and Longhi S 2013 Spectral and transport properties of time-periodic $\mathcal{PT}$-symmetric tight-binding lattices \emph{Phys. Rev. A} 87 022119
\bibitem{luo110} Luo X, Huang J, Zhong H, Qin X, Xie Q, Kivshar Y S and Lee C 2013 Pseudo-parity-time symmetry in optical systems \emph{Phys. Rev. Lett.} 110 243902
\bibitem{gong2015} Gong J and Wang Q 2015 Stabilizing non-Hermitian systems by periodic driving \emph{Phys. Rev. A} 91 042135
\bibitem{yang94} Yang B, Luo X, Hu Q and Yu X 2016 Exact control of parity-time symmetry in periodically modulated nonlinear optical couplers \emph{Phys. Rev. A} 94 043828
\bibitem{chit119} Chitsazi M, Li H, Ellis F M and Kottos T 2017 Experimental realization of Floquet $\mathcal{PT}$-symmetric systems \emph{Phys. Rev. Lett.} 119 093901
\bibitem{li2019} Li J, Harter A K, Liu J, de Melo L, Joglekar Y N and Luo L 2019 Observation of parity-time symmetry breaking transitions in a dissipative Floquet system of ultracold atoms \emph{Nat. Commun.} 10 855
\bibitem{xiao2017} Xiao L \emph{et al} 2017 Observation of topological edge states in parity-time-symmetric quantum walks \emph{Nat. Phys.} 13 1117
\bibitem{okolo} Okolowicz J, Ploszajczak M and Rotter I 2003 Dynamics of quantum systems embedded in a continuum \emph{Phys. Rep.} 374 271
\bibitem{bender80} Bender C M and Boettcher S 1998 Real spectra in non-Hermitian Hamiltonians having $\mathcal{PT}$ symmetry \emph{Phys. Rev. Lett.} 80 5243
\bibitem{bender89} Bender C M, Brody D C and Jones H F 2002 Complex extension of quantum mechanics \emph{Phys. Rev. Lett.} 89 270401
\bibitem{miri} Miri M A and Alu A 2019 Exceptional points in optics and photonics \emph{Science} 363 eaar7709
\bibitem{makris} Makris K G, El-Ganainy R, Christodoulides D N and Musslimani Z H 2008 Beam Dynamics in $\mathcal{PT}$ Symmetric Optical Lattices \emph{Phys. Rev. Lett.} 100 103904
\bibitem{longhi103} Longhi S 2009 Bloch Oscillations in Complex Crystals with $\mathcal{PT}$ Symmetry \emph{Phys. Rev. Lett.} 103 123601
\bibitem{feng333} Feng L, Ayache M, Huang J, Xu Y, Lu M, Chen Y, Fainman Y and Scherer A 2011 Nonreciprocal Light Propagation in Silicon Photonics \emph{Science} 333 729
\bibitem{regen488} Regensburger A, Bersch C, Miri M A, Onishchukov G, Christodoulides D N and Peschel U 2012 Parity-time synthetic photonic lattices \emph{Nature (London)} 488 167
\bibitem{hodaei} Hodaei H, Miri M A, Heinrich M, Christodoulides D N and Khajavikhan M 2014 Parity-time-symmetric microring lasers \emph{Science} 346 975
\bibitem{fleury} Fleury R, Sounas D and Alu A 2015 An invisible acoustic sensor based on parity-time symmetry \emph{Nat. Commun.} 6 5905
\bibitem{liu117} Liu Z, Zhang J, \"{O}zdemir S K, Peng B, Jing H, L\"{u} X, Li C, Yang L, Nori F and Liu Y 2016 Metrology with $\mathcal{PT}$ Symmetric Cavities: Enhanced Sensitivity near the $\mathcal{PT}$-Phase Transition \emph{Phys. Rev. Lett.} 117 110802
\bibitem{luoxb2023} Li Z, Hu X, Xiao J, Chen Y and Luo X 2023 Ratchet current in a $\mathcal{PT}$-symmetric Floquet quantum system with symmetric sinusoidal driving \emph{arXiv:} 2306.14095
\bibitem{kato} Kato Y K, Myers R C, Gossard A C and Awschalom D D 2004 Coherent spin manipulation without magnetic fields in strained semiconductors \emph{Science} 306 1910
\bibitem{bernevig} Bernevig B A, Hughes T L and Zhang S C 2006 Quantum Spin Hall Effect and Topological Phase Transition in HgTe Quantum Wells \emph{Science} 314 1757
\bibitem{lin471} Lin Y, Jim\'{e}nez-Garc\'{\i}a K and Spielman I 2011 Spin-orbit-coupled Bose-Einstein condensates \emph{Nature (London)} 471 83
\bibitem{wang109} Wang P, Yu Z, Fu Z, Miao J, Huang L, Chai S, Zhai H and Zhang J 2012 Spin-Orbit Coupled Degenerate Fermi Gases \emph{Phys. Rev. Lett.} 109 095301
\bibitem{cheuk109} Cheuk L, Sommer A, Hadzibabic Z, Yefsah T, Bakr W and Zwierlein M 2012 Spin-Injection Spectroscopy of a Spin-Orbit Coupled Fermi Gas \emph{Phys. Rev. Lett.} 109 095302
\bibitem{zhang109} Zhang J, Ji S, Chen Z, Zhang L, Du Z, Yan B, Pan G, Zhao B, Deng Y, Zhai H, Chen S and Pan J 2012 Collective Dipole Oscillation of a Spin-Orbit Coupled Bose-Einstein Condensate \emph{Phys. Rev. Lett.} 109 115301
\bibitem{huang12} Huang L, Meng Z, Wang P, Peng P, Zhang S, Chen L, Li D, Zhou Q and Zhang J 2016 Experimental realization of a two-dimensional synthetic spin-orbit coupling in ultracold Fermi gases \emph{Nat. Phys.} 12 540
\bibitem{wu354} Wu Z, Zhang L, Sun W, Xu X, Wang B, Ji S, Deng Y, Chen S, Liu X and Pan J 2016 Realization of Two-Dimensional Spin-Orbit Coupling for Bose-Einstein Condensates \emph{Science} 354 83
\bibitem{zhang128} Zhang S and Jo G 2019 Recent advances in spin-orbit coupled quantum gases \emph{J. Phys. Chem. Solids} 128 75
\bibitem{yu90} Yu Z and Xue J 2014 Selective coherent spin transportation in a spin-orbit-coupled bosonic junction \emph{Phys. Rev. A} 90 033618
\bibitem{zhang609} Zhang D, Fu L, Wang Z and Zhu S 2012 Josephson dynamics of a spin-orbit-coupled Bose-Einstein condensate in a double-well potential \emph{Phys. Rev. A} 85 043609
\bibitem{garcia89} Garcia-March M A, Mazzarella G, Dell'Anna L, Juli\'{a}-D\'{\i}az B, Salasnich L and Polls A 2014 Josephson physics of spin-orbit-coupled elongated Bose-Einstein condensates \emph{Phys. Rev. A} 89 063607
\bibitem{citro224} Citro R and Naddeo A 2015 Spin-orbit coupled Bose-Einstein condensates in a double well \emph{Eur. Phys. J. Spec. Top.} 224 503
\bibitem{kart97} Kartashov Y V, Konotop V V and Vysloukh V A 2018 Dynamical suppression of tunneling and spin switching of a spin-orbit-coupled atom in a double-well trap \emph{Phys. Rev. A} 97 063609
\bibitem{luo39} Li W, Yin H, Yi J, Luo Y, Xie X, Hai W and Luo Y 2022 Physics of manipulation of spin dynamics in a driven double well made
transparent \emph{Results Phys.} 39 105706
\bibitem{luo93} Luo Y, Lu G, Kong C and Hai W 2016 Controlling spin-dependent localization and directed transport in a bipartite lattice \emph{Phys. Rev. A} 93 043409
\bibitem{luo103} Luo X, Zeng Z, Guo Y, Yang B, Xiao J, Li L, Kong C and Chen A 2021 Controlling directed atomic motion and second-order tunneling of a spin-orbit-coupled atom in optical lattices \emph{Phys. Rev. A} 103 043315
\bibitem{luo22} Luo Y, Wang X, Luo Y, Zhou Z, Zeng Z and Luo X 2020 Controlling stable tunneling in a non-Hermitian spin-orbit coupled bosonic junction \emph{New J. Phys.} 22 093041
\bibitem{ren} Ren Z, Liu D, Zhao E, He C, Pak K, Li J and Jo G 2021 Chiral control of quantum states in non-Hermitian spin-orbit-coupled fermions \emph{arXiv:} 2106.04874
\bibitem{luo55} Tang J, Hu Z, Zeng Z, Xiao J, Li L, Chen Y, Chen A and Luo X 2022 Spin Josephson effects of spin-orbit-coupled Bose-Einstein condensates in a non-Hermitian double well \emph{J. Phys. B} 55 245301
\bibitem{luo2023} Luo Y, Wang X, Yi J, Li W, Xie X, Luo Z and Hai W 2023 Exact solutions for a spin-orbit coupled ultracold atom held in a driven double well \emph{J. Phys. A: Math. Theor.} 56 325302


\bibitem{zouml} Zou M, Lu G, Luo Y and Hai W 2020 Quantum transport and control of a classically chaotic open system \emph{Results Phys.} 17 103157
\bibitem{kreibich90} Kreibich M, Main J, Cartarius H and Wunner G 2014 Realizing $\mathcal{PT}$-symmetric non-Hermiticity with ultracold atoms and Hermitian multiwell potentials \emph{Phys. Rev. A} 90 033630
\bibitem{kreibich87} Kreibich M, Main J, Cartarius H and Wunner G 2013 Hermitian four-well potential as a realization of a $\mathcal{PT}$-symmetric system \emph{Phys. Rev. A} 87 051601(R)
\bibitem{shirley} Shirley J H 1965 Solution of the Schr\"{o}dinger equation with a Hamiltonian periodic in time \emph{Phys. Rev.} 138 B979
\bibitem{sambe} Sambe H 1973 Steady states and quasienergies of a quantum-mechanical system in an oscillating field \emph{Phys. Rev. A} 7 2203

\end{thebibliography}
\end{document}